\documentstyle[psfig]{mn}

\newif\ifAMStwofonts

\def\lapp{\ifmmode\stackrel{<}{_{\sim}}\else$\stackrel{<}{_{\sim}}$\fi}
\def\gapp{\ifmmode\stackrel{>}{_{\sim}}\else$\stackrel{>}{_{\sim}}$\fi}

\def\psr{PSR~B0906$-$49}

\title[High-precision geometry of a double-pole pulsar]
{High-precision geometry of a double-pole pulsar}
\author[Kramer \& Johnston]
{Michael Kramer$^1$ \& Simon Johnston$^2$\\
$^1$ University of Manchester, Jodrell Bank Centre for
Astrophysics Alan-Turing Building, Oxford Road, Manchester M13 9PL, UK\\
$^2$Australia Telescope National Facility, CSIRO, P.O. Box 76, 
Epping, NSW 1710, Australia. \\
}
\date{\today}
\pagerange{\pageref{firstpage}--\pageref{lastpage}}
\pubyear{2008}
\begin{document}
\maketitle
\label{firstpage}

\begin{abstract}
%
%
%
  High time resolution observations of \psr\ (or PSR J0908$-$4913) over a wide
  range of frequencies have enabled us to determine the geometry and beam
  shape of the pulsar.  We have used the position angle traverse to determine
  highly-constrained solutions to the rotating vector model which show
  conclusively that \psr\ is an orthogonal rotator. The accuracy obtained in
  measuring the geometry is unprecedented. This may allow tests of high-energy
  emission models, should the pulsar be detected with GLAST. Although the
  impact parameter, $\beta$, appears to be frequency dependent, we have shown
  that this is due to the effect of interstellar scattering.  As a result,
  this pulsar provides some of the strongest evidence yet that the position
  angle swing is indeed related to a geometrical origin, at least for
  non-recycled pulsars. We show that the beam structures of the main pulse and
  interpulse in \psr\ are remarkably similar. The emission comes from a height
  of $\sim$230~km and is consistent with originating in a patchy cone located
  about half way to the last open field lines.  The rotation axis and
  direction of motion of the pulsar appear to be aligned.
\end{abstract}

\begin{keywords}
pulsars:general -- pulsars:individual:PSR J0908$-$4913 -- stars:neutron
\end{keywords}

\section{Introduction}

Pulsars are highly magnetized, rotating neutron stars which emit beams of
radio emission at a low height above the star's magnetic poles.  The emission
is confined to the open field lines and the beam only sweeps out a small
fraction of the celestial sphere. For this reason we see emission from only
one pole in the vast majority of pulsars.  In fortunate geometrical
circumstances, however, the second pole should also be observable as a radio
emitting site, if the spin axis is almost orthogonally orientated to the
line-of-sight to Earth. Such pulsars can be identified by the detection of an
``interpulse'' separated by $\sim 180^\circ$ in pulse longitude from the main
pulse.  Less than 30 pulsars have been identified to show interpulses
\cite{rib08,wj08a} from the known population of $\sim$1800 pulsars.  Not all
interpulses may arise from the second magnetic pole, but some of them can also
be interpreted within a geometry of an aligned rotator where the observer's
line-of-sight never, or hardly, leaves the emission cone during the pulse
period (e.g.~Manchester \& Lyne 1977).\nocite{ml77}

The geometry of a pulsar may be determined from observations of their
polarization which has been a major industry since shortly after
their discovery. In the more than 40 years since, a few key points are well
established (see e.g.~Lorimer \& Kramer 2005).\nocite{lk05} 
Pulsars are essentially rotating dipoles and the super-strong
magnetic field naturally implies a high degree of linear polarization with the
position angle (PA) of the radiation parallel (or perpendicular) to the
magnetic field direction. Therefore, as the dipole field sweeps past the
observer once per rotation of the pulsar, the observer sees a characteristic S
shaped sweep of the position angle. Known as the rotating vector model (RVM),
it provides a simple mapping between the relevant geometrical angles and the
position angle variations \cite{rc69a}.  The PA as a function of pulse
longitude, $\phi$, can be expressed as
\begin{equation}
{\rm PA} = {\rm PA}_{0} +
{\rm arctan} \left( \frac{{\rm sin}\alpha
 \, {\rm sin}(\phi - \phi_0)}{{\rm sin}\zeta
 \, {\rm cos}\alpha - {\rm cos}\zeta
 \, {\rm sin}\alpha \, {\rm cos}(\phi - \phi_0)} \right)
\end{equation}
Here, $\alpha$ is the angle between the rotation axis and the magnetic axis,
and $\zeta=\alpha+\beta$, with $\beta$ being the angle at closest approach of
the line of sight to the magnetic axis.  $\phi_0$ is the corresponding pulse
longitude at which the PA is then PA$_{0}$.  Note that we follow the
observer's sign convention for the PA, which differs from that introduced by
Damour \& Taylor (1992\nocite{dt92}, see also Everett \& Weisberg
2001\nocite{ew01}).

The RVM has been used extensively to derive geometrical angles in a number of
pulsars \cite{ran83,lm88}. Unfortunately there are many factors which mean
that $\alpha$ and $\beta$ are often subject to large uncertainties. These
include a covariance between $\alpha$ and $\beta$ which cannot be broken over
the usually small longitude range over which pulsars emit (e.g.~von
Hoensbroech \& Xilouris 1997\nocite{hx97a}, Everett \& Weisberg
2001\nocite{ew01}), but also orthogonal modes and propagation effects in the
pulsar magnetosphere \cite{ew01}. For these reasons multi-frequency studies of
a given pulsar rarely give similar values for $\alpha$ and $\beta$ as the
uncertainties are large (e.g. Mitra \& Li 2004\nocite{ml04}).

Some information about the geometry can also be obtained from the evolution of
pulse profiles with frequency.  Observational data show that low frequency
pulse profiles tend to be wider in extent than high frequency profiles
\cite{tho91a} and the explanation for this is that radio emission originates
at different heights in diverging dipolar magnetic field lines.  In this
model, high frequency radio emission originates from closer to the stellar
surface than low frequency emission with emission heights ranging from several
tens to several hundreds of km \cite{mr02a}. Recent evidence tends to suggest
that, at a given frequency, the emission height varies across the polar cap
with lower emission heights over the magnetic axis \cite{gg03}.  Also,
observations indicate that low-frequency profiles are often dominated by
strong central components whereas high-frequency profiles show strong outrider
components in many cases. This has been interpreted as indicating a different
spectral index for core and cone emission \cite{ran83}, while geometrical
factors may be at least partly responsible \cite{kwj+94}.

When considering emission heights, it is intriguing to also consider the
observational consequences of relativistic effects in the magnetosphere.
Blaskiewicz, Cordes \& Wasserman (1991)\nocite{bcw91} pointed out that, to
first order, the main effect is that the PA swing is delayed with respect to
the total intensity profile and the magnitude of this effect is directly
propotional to the emission height. This effect should therefore be frequency
dependent if different frequencies originate at different heights. The studies
by Blaskiewicz et al.~(1991), von Hoensbroech \& Xilouris (1997) and Mitra \&
Li (2004)\nocite{ml04} showed evidence for this effect in a sample of pulsars.

In this paper, we present observations and their analysis of \psr\, a 100-ms
pulsar discovered by D'Amico et al.~(1988)\nocite{dmd+88} which allows us, as
it turns out, to study our described understanding of pulsar emission.
Initial observations showed that the profile consisted of two widely separated
components.  Polarization observations at 0.66 and 1.6~GHz were carried out by
Qiao et al.~(1995)\nocite{qmlg95} and Wu et al.~(1993)\nocite{wmlq93}
respectively. They showed that the profile is virtually 100 per cent linearly
polarized with an indication of a small fraction of circular polarization. The
time resolution of the observations was low and each component was largely
featureless. Wu et al. (1993)\nocite{wmlq93} attempted an RVM fit to the PA
swing and determined that $\alpha=58\degr$ and $\beta=30\degr$.  This fit
indicated that the emission arises from a single pole and supported the idea
that young pulsars had wide profiles \cite{man96}.

In a survey looking for pulsar wind nebulae, Gaensler et
al.~(1998)\nocite{gsfj98} discovered an unusual looking nebula around
\psr. The head of the nebula is clearly resolved and a bow-shock indicates
that the direction of motion of the pulsar is at an angle of $\sim$315\degr\
(measured as north through east).  Polarization properties of the pulse
profile can give an indication of the position angle of the pulsar's rotation
axis and we will investigate if the correlation between the velocity and
rotation axis seen by Johnston et al. (2005)\nocite{jhv+05} is also evident in
this pulsar, and whether it confirms the direction of motion suggested by
Gaensler et al.~(1998).

The organisation of the paper is as follows. In Section 2 we briefly discuss
multi-frequency observations of \psr\ as part of a bigger campaign to measure
polarization in southern pulsars generally.  In Section 3 we show how the
profile of \psr\ evolves with frequency and present our RVM fits, and
re-examine in particular the question of whether this pulsar shows emission
from a single, wide pole or from two poles. We discuss the implications of the
RVM fits in Section 4 and derive a model for the beam of \psr.

\begin{figure}
\centerline{\psfig{figure=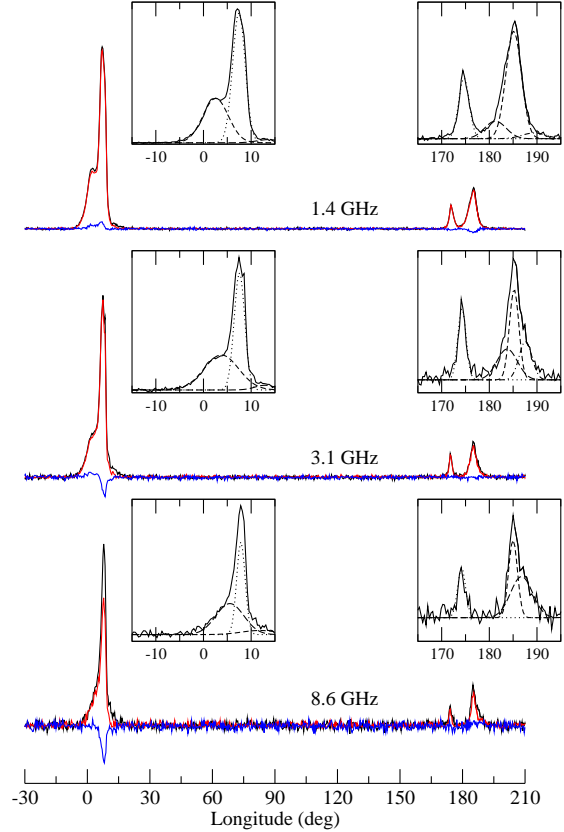,angle=0,width=9.5cm}}
\caption{\label{fig:profiles}
Polarization profiles of \psr\ obtained at three frequencies.
While the interpulse shows hardly any frequency evolution, the
main pulse exhibits an unusual behaviour (see text). The circular
polarisation changes sign between the low and high frequencies.
The insets show a Gaussian decomposition of the main and interpulse.
}
\end{figure}

\section{Observations}

All observations were carried out using the Parkes 64-m radio telescope
located in central New South Wales, Australia.  \psr\ was observed on a number
of occasions at a range of frequencies over the period 2004 December to 2007
July.  The observing bands were centred on frequencies of 1.37, 3.10 and 8.6
GHz.  The bandwidths used were 256, 1024 and 512 MHz respectively.  In all
cases, the backend used was the wideband correlator specifically designed for
high-time-resolution polarimetric observations of pulsars. At all frequencies
we sampled the data with 1024 phase bins across the pulse period. The
correlator performs on-line folding and de-dispersion at the topocentric
period and dispersion measure.  Before each observation of the pulsar, a
pulsed calibration signal was observed to allow correction of the gain and
phase between the two feed probes.  At each frequency observations were made
of the flux calibrator Hydra A to provide accurate flux densities.

Data reduction was carried out using the PSRCHIVE software package
\cite{hvm04}.  The final product was a fully polarimetric- and flux-calibrated
pulse profile at each frequency. We are able to obtain absolute position
angles for the linear polarization using techniques described in Johnston et
al.~(2005)\nocite{jhv+05}.  From multi-frequency observations made in 2005
July, the rotation measure of \psr\ was determined to be 13.0$\pm$0.2
rad~m$^{-2}$ (cf 10.0$\pm$1.6 rad~m$^{-2}$ in Qiao et al. 1995).

\section{Results}

\subsection{Profile evolution}

Figure \ref{fig:profiles} shows the polarization profiles of \psr\ obtained at
1.4, 3.1 and 8.6~GHz. Zero longitude coincides with the fiducial point derived
from beam considerations discussed in Sections~\ref{sec:abb} and
\ref{sec:beam}.  Values for $\phi_0$ quoted in Table~\ref{tab:paresults} were
measured relative to this longitude.

The higher time and frequency resolution of our data compared to that of Wu et
al. (1993) allows us to distinguish features in the profile not previously
seen.  The main pulse (MP) consists of a small leading component blended with
a stronger, narrower trailing component. In contrast, the interpulse (IP) is a
separated double profile with the trailing component somewhat stronger than
the leading component. Johnston \& Weisberg (2006)\nocite{jw06} have pointed
out that a double-pulse structure with the trailing component dominating
appears typical of young pulsars generally and both the MP and IP show this
behaviour.  The MP-IP separation remains essentially constant at all
frequencies and the intensity ratio of MP and IP remains unchanged, indicating
similar radio spectra.

The IP shows hardly any frequency evolution, although the
central part between the prominent components becomes noticeably weaker,
leading to sharper inner edges of the components. This behaviour is
typical for many pulsars \cite{lm88}.  In contrast, the MP
undergoes clear changes with frequencies. The dominant component
remains the strongest at all frequencies, while the outer
components (in particular the leading one) become weaker at higher
frequencies.

\subsection{Polarization and its evolution}

Typical of young pulsars, the profile of \psr\ (both MP and IP) is 
virtually 100 per cent linearly polarized. Only at 8.4~GHz does the
polarization appear to decrease with both MP and IP having a fractional
polarization near 70 per cent.

At 1.4~GHz the MP shows positive circular polarization in the first two
components with a hint of negative circular in the trailing
component. The value of $|V|$ is 4.1 per cent.  In the IP, there is no
circular polarization in the leading component but about 9 per cent of
negative circular polarization in the trailing component.  At 3.1~GHz the
polarization has changed sign in the dominant component of the MP - there
is little or no circular polarization at the leading component but about 9 per
cent of negative circular polarization in the trailing component the peak
of which is offset from the total intensity peak.  In the IP the circular
polarization remains negative in the trailing component with a value of
about 7 per cent.  At 8.6~GHz the circular polarization has increased in
the main component of the MP.  There is now 15 per cent of negative circular
polarization and the peak of the circular is aligned with the total intensity
peak.  In the IP, the circular polarization is rather low with an upper limit
of about 6 per cent but appears to be positive in sign. In other words, it
seems that the MP and IP show a mirror-symmetry in changing handedness of
circular polarisation. We note that in the low frequency profiles of Qiao et
al. (1995), there appears to be little circular polarization in the
interpulse but significant positive circular in the main pulse.

The picture where circular polarization is stronger in the trailing
component in young pulsars was pointed out by Johnston \& Weisberg (2006)
but this is one of the few pulsars in which the sign of $V$ changes as a 
function of frequency. In the compilation of Han et al. (1998)\nocite{hmxq98}
only PSR~B1240$-$64 shows unambiguous evidence for a sign change
from low to high frequencies.

\begin{table*}
\caption{\label{tab:paresults}
Geometry determined from RVM fits at three frequencies. While the left
columns show results obtained by freely fitting for all angles and
the resultant reduced $\chi^2$, the
last column lists $\beta$ obtained after keeping $\alpha$ fixed at an
angle of 96\degr.  Quoted uncertainties were estimated using Monte-Carlo
simulations.}
\begin{tabular}{lcccccc}
\hline
\hline
$\nu$ (GHz) & $\alpha$ (deg) & $\beta$ (deg) & $\zeta$ (deg) &
$\phi_0$ (deg) & $\chi^2_{\rm red}$ & $\beta^\ast$ (deg)  \\
\hline
1.4 & $96.60\pm0.03$ &  $-8.1\pm0.1$ & $88.5\pm0.1$ &  $11.31\pm0.05$ & 0.98 & $-8.47\pm0.08$ \\
3.0 & $96.22\pm0.07$ &  $-6.4\pm0.2$ & $89.2\pm0.2$ & $11.35\pm0.06$  & 1.30 & $-6.4\pm0.2$ \\
8.4 & $96.1\pm0.4$ &  $-5.9\pm0.6$ &  $90.2\pm0.6$ & $10.8\pm0.3$  & 0.82& $-6.0\pm0.5$\\
\hline
\end{tabular}
\end{table*}
\begin{figure}
\centerline{\psfig{figure=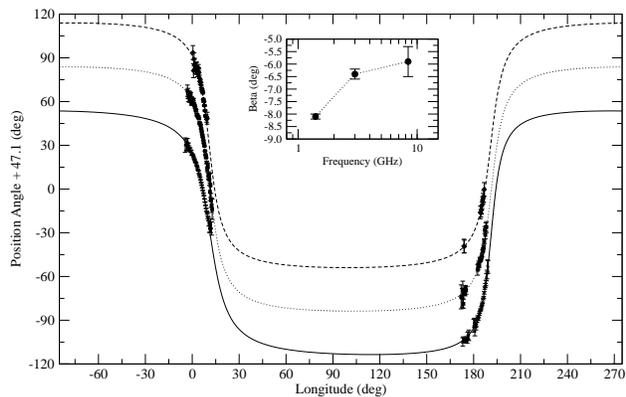,angle=-90,width=9.5cm}}
\caption{\label{fig:pafit}
Position angle of the linearly polarized emission and
RVM fits. The 1.4-GHz
data and their fit (solid line) and the 8.4-GHz data and
their fit (dashed line) are offset by $-30$ and $+30$ deg,
respectively, to the 3.0-GHz data (dotted line) for clarity.}
\end{figure}

\subsection{Rotating-Vector-Model fits}

Figure \ref{fig:pafit} shows the measured values of the PAs of the linear
polarisation, $\Psi$, and fits of the RVM to those.  In contrast to the large
majority of pulsars where the small duty cycle prevents a reliable
determination of the RVM parameters \cite{ew01}, the highly polarized IP of
\psr\ allows a very precise measurement of the magnetic inclination angle
$\alpha$, the impact parameter $\beta$ (or alternatively, the line-of-sight
angle $\zeta=\alpha+\beta$), and the offset angles $\Psi_0$ and $\phi_0$,
determining the symmetry positions of the fitted curve in PA and longitude,
respectively. Table \ref{tab:paresults} lists the results obtained for each
frequency. We obtain excellent fits with a reduced $\chi^2$ of close to unity
at all three frequencies, clearly indicating an orthogonal geometry. This is
in contrast to the results of Wu et al.~(1993),\nocite{wmlq93} who obtained
$\alpha$=58\degr. Inspecting their Fig.~4, we note that our PA values for the
IP are offset from theirs by 180\degr\ (an inherent ambiguity to measured
PAs). Performing a fit to our PA values in such an arrangement is possible but
results in much worse fits ($\chi_{\rm red}^2>200$), so that we believe that
our solution is the correct one. Indeed, a free fit to all parameters results
in consistent values at all frequencies, in particular, in $\alpha$ close to
96\degr. Given the problems often encountered in obtaining reliable RVM fits,
this is an extraordinary result. The results are so precise that they reveal a
surprising, unexpected decrease in the magnitude of the impact angle $\beta$
with increasing frequency. We confirm this observation by keeping $\alpha$
fixed in subsequent fits (see Table~\ref{tab:paresults}) and a visiual
inspection of Figure \ref{fig:pafit} which indeed shows a steeping of the PA
curve at the inflexion point $\phi_0$. We see the same trend, albeit with much
larger uncertainties, by fitting the MP and IP separately. Such fits can also
be used to determine the $\phi_0$ independently for MP and IP. Doing this by
keeping $\alpha$ fixed to $96\degr$ (see below) to increase the accuracy, we
measure an MP-IP separation of $\phi_0$(MP)$-\phi_0$(IP)$=180.0\pm0.2\degr$.

For reason discussed in detail in Section~\ref{scattering}, we consider the
solution obtained at the highest frequency as the correct intrinsic
geometry. We therefore come to the conclusion that $\alpha_{\rm MP}$=96\degr,
$\beta_{\rm MP}$=$-$6\degr, $\alpha_{\rm IP}$=84\degr, $\beta_{\rm
  IP}$=6\degr.  This implies that we are seeing the MP and IP at the same
impact angle on opposite sides of the pole. This is unusual in itself, compare
for example the situation in PSR~B1702$-$19, where the impact angles for the
MP and IP are quite different \cite{wws07}. Finally, the inflexion point of
the RVM occurs {\it later} than either the profile midpoint or the profile
peaks.

\subsection{Rotation axis and proper motion direction}

As outlined in the introduction, Gaensler et al.~(1998)\nocite{gsfj98}
detected a bow-shock nebula around \psr, with the inference that the direction
of motion was $\sim$315\degr. The RVM fits described above show that the
rotation axis points towards 48\degr or 228\degr. There is therefore an offset
of 87\degr\ between the proper motion direction and the rotation axis. We have
shown elsewhere \cite{jhv+05,jkk+07} that the probable implication of this is
that the proper motion vector and rotation axis are likely to be aligned, with
the pulsar emitting linear polarisaion in a mode with the plane of
polarisation orthogonal to the magnetic field direction. This is consistent
with our earlier results obtained from different samples of pulsars, and also
confirms the interpretation of the bow-shock nebula by Gaensler et al.~(1998).

\begin{figure}
\centerline{\psfig{figure=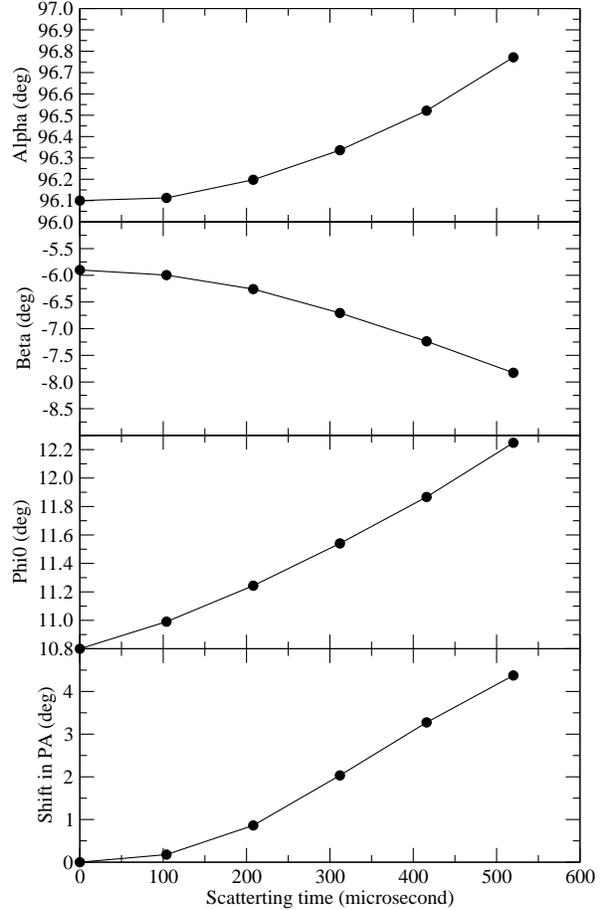,angle=0,width=9.5cm}}
\caption{\label{fig:sim}
Results of simulations investigating the impact of weak interstellar
scattering on the observed PA swing and the derived geometry. Systematic
changes in the determined values are clearly visible. Values for zero
scattering correspond to those observed at 8.4 GHz.}
\end{figure}

\section{Discussion}

\subsection{Frequency-dependent geometry?}
\label{scattering}

The geometrical interpretation of the RVM implies that the determined values
for $\alpha$ and $\beta$ should be independent of frequency.  Therefore, our
results of a changing $\beta$ and $\phi_0$ reported in Figure \ref{fig:pafit}
and Table~\ref{tab:paresults} are somewhat surprising. Usually, the geometry
of a pulsar cannot be determined with such high precision as is possible here,
so that this effect may not have been noticed for other sources before.
However, rather than questioning the geometrical interpretation of the RVM,
other explanations may apply. Indeed, as we show in the following, the
apparent change can be explained by an undetected scattering of the pulse
profile at lower frequencies.

We investigated the impact of interstellar scattering on the observed PA swing
in a similar fashion to that presented in the research note of Li \& Han
(2003)\nocite{lh03}.  We confirm their results that large scattering leads to
flat PA curves over the longitudes of the scattering tail. Such obvious
effects of scattering are, however, not visible in the data for \psr\ 
presented here. Hence, we studied the impact of much smaller
scattering times. For this we used a noise-free template constructed from the
determined Gaussian decomposition of the observed profile. We assigned a
constant degree of linear polarisation and computed Stokes $Q$ and $U$ from
the RVM fits to the 8.4 GHz data as shown in Table~\ref{tab:paresults}. We
then convolved the Stokes $I$, $Q$ and $U$ with a scattering tail as caused by
a thin screen in the interstellar medium located half-way between the pulsar
and the observer (e.g.~Williamson 1972)\nocite{wil72}. The resulting PA swing
was computed from the scattered Stokes $Q$ and $U$ intensities and subjected
to the same RVM fitting process as the real data. This process was repeated
for scattering times in the range from zero to five phase bins, yielding the
results shown in Figure~\ref{fig:sim}.

The general agreement between the simulations and the observations is
remarkable. While the predicted change in $\phi_0$ is somewhat too large, the
behaviour of the measured $\alpha$ and $\beta$ is well reproduced. This
confirms that, based on the polarization profile observed at 8.4 GHz, the PA
swings (and the derived values for $\alpha$ and $\beta$) at 3.0 and 1.4 GHz
can be well explained by a scattering of the profile with scattering times of
only about 200$\mu$s and 500$\mu$s, respectively. These scattering times are
so small, that the corresponding total power profiles are hardly
distinguishable from the unscattered profile in our simulations. Future
high-resolution low-frequency studies should enable us to trace the frequency
evolution of these scattering effects.

While the geometry derived from the RVM is therefore indeed independent of
radio frequency, we note that the simulations also produce a shift in absolute
PA that does depend on observing frequency.
Even though this shift is only 4\degr\ over a
frequency range from 1.4 to 8.4 GHz, we point out that this effect might
affect rotation measures determined from RVM fits over widely separated
frequencies. However, the precision to which the geometry
can be determined for \psr\ is unrivalled, so that in practice this effect may
be small compared to other sources of uncertainties.

\subsection{Aberration effects}
\label{sec:abb}

The precision obtained in fitting the RVM model to the observed data, promises
to allow us to study aberration effects in pulsar radio emission. A
first-order relativistic treatment of the RVM leads to the expectation that
the inflexion point of the PA swing, $\phi_0$, should be delayed with respect
to the profile midpoint by an amount $\Delta t_{PA} = 4\times r_{em}/c$ where
$r_{em}$ is the emission height \cite{bcw91,ha01,drh04,dyk08}.

The main difficulty in showing whether aberration is occuring or not is the
determination of the location of the pulsar's fiducial point. For pulsars with
symmetrical pulse profiles, one generally locates the fiducial point in the
profile centre but for more complex pulse profiles the situation is much less
clear although one can also use the profile's frequency evolution as a help.
In the case of \psr, the MP is rather asymmetric (and becomes more so at
higher frequencies). For the IP, the profile is much more symmetrical (and
remains so at high frequencies), suggesting that the IP midpoint should be
separated by 180\degr\ from the fiducial point of the MP. This is indeed how we
chose the zero-degree reference point for the pulse longitude in
Fig~\ref{fig:profiles}, even though it is clearly not centred on the MP
profile. We discuss this issue further in the
next section.

\subsection{Beam Description}
\label{sec:beam}

We can draw all the strands of the above arguments together to come up with a
plausible beam model for \psr\ for the following reasons.  We believe the best
location for the fiducial point of the MP and IP is at 0\degr\ and 180\degr as
shown in Fig~\ref{fig:profiles}.  First, this is the midpoint of the
symmetrical IP profile. Secondly, there is a hint of a central component at
this location in the IP at 1.4~GHz and this has a steeper spectral index than
the outer components in accordance with the general rule for central
components. This implies that the leading component in the MP is more
centrally located (and also has a steeper spectral index) and we see emission
at the trailing edge but not at the leading edge of the emission cone.

The offset between our purported profile fiducial point and the inflexion
point as determined by the RVM is therefore some 11\degr. This value can be
used to determine an emission height of $\sim$230~km and the emission from
both the MP and IP must originate from approximately the same height.  In
turn, the emission height can be used to determine the half-opening angle of
the open fieldline region, $\rho$, which must then be $\sim$18\degr. This is
about a factor of two larger than the measured half-width of the profile which
is about 9\degr\ for both the MP and IP (although we do not see the leading
emission in the MP). Interestingly, the derived value of $\rho$ for \psr\ is
very close to that given by the relationship $\rho=6.3/\sqrt{P}$ with the
pulse period $P$ measured in seconds (e.g.~Kramer et al. 1994\nocite{kwj+94}),
determined empirically from a large sample of pulsars.
\begin{figure}
\centerline{\psfig{figure=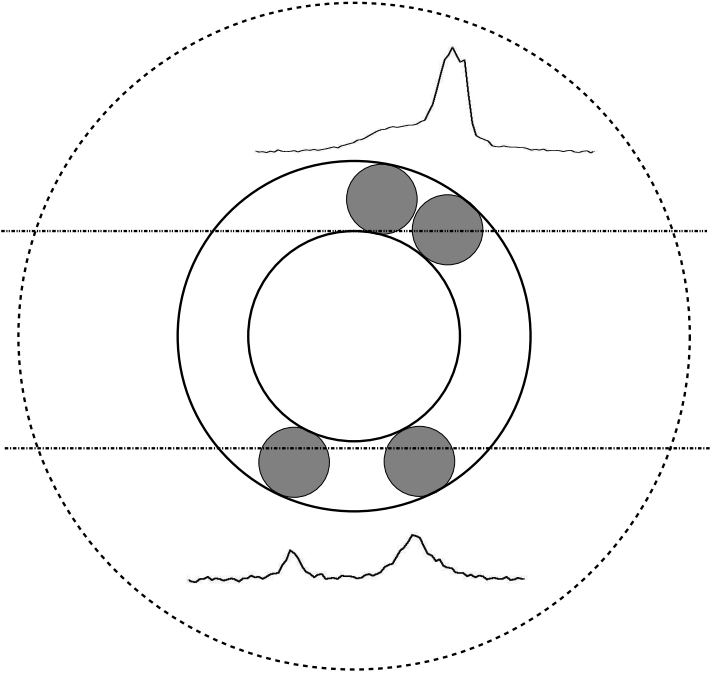,angle=0,width=9.5cm}}
\caption{\label{fig:geo} A possible configuration for the beam of \psr. The
  active regions in the magnetosphere appear to fall into the same cone for
  both the MP and the IP and so we have shown only one cone here. The outer
  dotted circle represents the region bounded by the last open field
  lines. The active emission cone is shown by the solid circles; its location
  and width is dictated by the width of the pulse profiles.  The line of sight
  for the MP traverse is the top solid line; it cuts two active patches near
  the centre and on the trailing edge of the cone.  The line of sight for the
  IP traverse (bottom solid line) also cuts through two active patches but
  they are more symmetrically placed about the beam centre. The resultant
  pulse profiles (MP and IP) are also shown.}
\end{figure}

In Figure~\ref{fig:geo} we have drawn a sketch of a possible configuration of
the beam for both the MP and IP. The emission appears to arise from a cone of
thickness $\sim$3\degr\ which is located about half way out to the last open field
lines. The emission cone is not fully illuminated but rather consists of
active patches. Both the MP and the IP sight lines appear to cross two such
patches. What is remarkable is that the active regions in the magnetosphere
near both poles happen to lie on the same cone.

\section{Conclusions}

High time resolution observations over a range of frequencies for \psr\ have
enabled us to determine the geometry and beam shape of the pulsar.  Uniquely
amongst pulsars, even amongst those with interpulses, we have been able to
determine highly constrained RVM fits which show conclusively that \psr\ is an
orthogonal rotator. We have shown that the value of $\beta$ appears to be
frequency dependent; however careful modelling shows that this appears to be
entirely due to the effect of interstellar scattering of the profiles. While
other factors usually dominate the uncertainties in the geometry derived from
RVM fits, it should be considered in particular for low-frequency observations
or observations of short-period pulsars. It will also be relevant for future
high-sensitivity detections and studies of interpulse sources with the
Square-Kilometre-Array \cite{ckl+04}.

Interestingly, the relativlely high ratio of spin-down luminosity $\dot{E}$ to
distance $d$ as measured by $\dot{E}/d^2 = 1.1\times 10^{34}$ erg s$^{-1}$
kpc$^{-2}$ makes this pulsar potentially detectable with the Gamma-Ray Large
Area Space Telescope (GLAST). This would be an exciting opportunity to test
competing $\gamma$-ray emission models, given the very well determined
geometry of this pulsar. For instance, outer gap models (e.g.~ Chiang \&
Romani 1994)\nocite{cr94} predict that orthogonal rotators should be
$\gamma$-ray bright as we look directly into the null-charge surface, i.e.~the
bottom part of the outer gap (Yadigaroglu \& Romani 1995).  \nocite{yr95}

We have shown that the total intensity structure of the MP and the IP in \psr\
have similarities. While the degree of linear polarisation is very large
in both MP and IP, the circular polarisation changes handedness with
increasing frequency with an intriguing mirror-symmetry between the MP and IP.
The emission comes from a height of $\sim$230~km in both and is consistent
with an origin in a cone with a radius of only one half of the beam opening
angle. The cone is patchy as we do not detect the leading emission from the
MP.

In summary, this pulsar represents a show-case for many features of pulsar
radio emission, combined in a single, remarkable source. But most of all, this
source provides conclusive evidence for the geometrical interpretation of the
PA swing within the RVM. Many pulsars, however, in particular recycled
pulsars, show large deviations from a simple RVM swing in which cases the
geometrical interpretation has to be questioned. In this case, however, the
evidence is incontrovertible.

\section*{Acknowledgments}
The Australia Telescope is funded by the Commonwealth of Australia for
operation as a National Facility managed by the CSIRO. We thank Andrew Lyne
and the referee Jarek Dyks for useful and stimulating comments on the
manuscript.


\label{lastpage}
\end{document}